\documentclass[pra,onecolumn,showkeys,superscriptaddress]{revtex4-2}

\usepackage[dvipsnames]{xcolor}
\usepackage{graphicx}
\usepackage{amsmath}
\usepackage{amsthm}
\usepackage{amssymb}
\usepackage{times}
\usepackage{braket}

\usepackage[T1]{fontenc}

\newtheorem{proposition}{Proposition}

\theoremstyle{definition}

\newtheorem{example}{Example}

\begin{document}

\title{Superior resilience of non-Gaussian entanglement against local Gaussian noises}

\author{Sergey Filippov}
\affiliation{Algorithmiq Ltd, Kanavakatu 3 C, 00160 Helsinki, Finland}

\author{Alena Termanova}
\affiliation{Terra Quantum AG, Kornhausstrasse 25, 9000 St. Gallen, Switzerland}

\begin{abstract}
Entanglement distribution task encounters a problem of how the initial entangled state should be prepared in order to remain entangled the longest possible time when subjected to local noises. In the realm of continuous-variable states and local Gaussian channels it is tempting to assume that the optimal initial state with the most robust entanglement is Gaussian too; however, this is not the case. Here we prove that specific non-Gaussian two-mode states remain entangled under the effect of deterministic local attenuation or amplification (Gaussian channels with the attenuation factor/power gain $\kappa_i$ and the noise parameter $\mu_i$ for modes $i=1,2$) whenever $\kappa_1 \mu_2^2 + \kappa_2 \mu_1^2 < \frac{1}{4}(\kappa_1 + \kappa_2) (1 + \kappa_1 \kappa_2)$, which is a strictly larger area of parameters as compared to where Gaussian entanglement is able to tolerate noise. These results shift the ``Gaussian world'' paradigm in quantum information science (within which solutions to optimization problems involving Gaussian channels are supposed to be attained at Gaussian states).
\end{abstract}


\maketitle

\section{Introduction}

Quantum entanglement is a feature of genuinely quantum correlations that underly many interesting physical phenomena and are a primary resource in quantum information theory enabling advantageous protocols of information processing and transmission~\cite{horodecki-2009}. A typical scenario to share this resource among distant parties is to prepare an entangled state in a laboratory and send entangled parts of the state toward the recipients via quantum communication lines~\cite{hanzo-2020}. Upon receiving the respective components of the entangled state the recipients can utilize them in entanglement-enabled protocols such as device-independent quantum key distribution~\cite{acin-2007,zhang-2022,nadlinger-2022} or further convert these entangled degrees of freedom into other degrees of freedom at their disposal, thus creating an entanglement among particles or modes that have never interacted before (entanglement swapping)~\cite{jin-sasaki-2015}. The crucial practical challenge in this scenario is that the quantum communication lines are far from being ideal and introduce noise that degrades entanglement. The noise is local because the signals propagate to recipients via different communication lines whose environments do not interact with each other. There are some known techniques on how to locally compensate the negative effect of noise and distill some number of highly entangled states from many poorly entangled ones~\cite{rozpedek-2018}. However, if the noise destroys entanglement completely so that the recipients actually receive a fully separable quantum state, no local strategy is able to cure this state and make it entangled. Therefore, there exists a fundamental limitation on the noise intensity exceeding which no entanglement can survive --- a manifistation of the entanglement annihilation phenomenon~\cite{moravcikova-ziman-2010}.

The maximally permissible noise level for entanglement preservation apparently depends on the physical nature of the quantum states sent, the degrees of freedom used, and the nature of the noise that affects those degrees of freedom. In the case of two recipients $A$ and $B$ a local noise is mathematically described by a completely positive trace preserving map $\Phi_1^A \otimes \Phi_2^B$. The quantum channels $\Phi_1$ and $\Phi_2$ can differ in general as the noises in the communication lines do not have to be identical due to, e.g., a different length or a different physical environment, say, atmosphere~\cite{bohmann-2016}, optical fiber~\cite{wengerowsky-2019,krutyanskiy-2019}, photonic chip~\cite{chen-2021} or even a quantum memory cell, which can also be considered as a quantum channel. Historically, the first fully studied case of entanglement dynamics is the case of one-sided noise for which $\Phi_1 = {\rm Id}$, the identity transformation. In this case the local map ${\rm Id}^A \otimes \Phi_2^B$ degrades entanglement of any state if and only if $\Phi_2$ has a measure-and-prepare (entanglement-breaking) structure~\cite{holevo-1998,horodecki-2003,konrad-2008}. For the one-sided noise, the optimal initial state for entanglement preservation is maximally entangled. The case of a two-sided noise $\Phi_1^A \otimes \Phi_2^B$ is more complicated to study even for simple systems (such as qubits) and relatively simple noise models (such as $T_1$-$T_2$ qubit decoherence) because the optimal initial state for entanglement preservation is not known a priori. The entanglement annihilation parameters for a local two-qubit depolarizing noise were found in Ref.~\cite{tiersch-2009}. Local two-qubit unital noises were studied in Ref.~\cite{filippov-rybar-ziman-2012} and then generalized with the help of the quantum Sinkhorn theorem to the case of an arbitrary two-qubit local noise $\Phi_1^A \otimes \Phi_2^B$~\cite{ffk-2018}. The case of trace decreasing qubit maps (corresponding to the loss of particles) was analyzed in Ref.~\cite{filippov-2021}. The higher dimensional systems (qudits) were considered in Refs.~\cite{filippov-ziman-2013,filippov-2014,lami-2016}, with a full characterization of entanglement annihilation parameters being obtained in the case of local qudit-depolarizing noises~\cite{lami-2016}.

The continuous-variable quantum states of electromagnetic radiation (in contrast to discrete-variable polarization states) naturally incorporate the loss of photons within their description and encode quantum information in the field amplitude. The vacuum, coherent, squeezed, and thermal states are typical examples of Gaussian continuous-variable quantum states whose characteristic function is Gaussian~\cite{dodonov-2002,adesso-2014,serafini-2017}. These states are conventionally sensed by means of the homodyne and heterodyne measurements, which in turn are Gaussian quantum measurements~\cite{holevo-2019}. The losses in quantum communication lines (potentially accompanied by an admixture of extra noise) are described by a quantum attenuator channel $\Phi(\kappa,\mu)$, where $\kappa \in [0,1)$ is the intensity attenuation factor and $\mu \geq \frac{1}{2}(1 - \kappa)$ is the noise parameter precisely defined in terms of characteristic functions in Section~\ref{section-channels}. Similarly, a deterministic phase-insensitive amplification of optical signal is described by a quantum linear amplifier $\Phi(\kappa,\mu)$, where $\kappa \in (1,\infty)$ is the power gain and $\mu \geq \frac{1}{2}(\kappa - 1)$ is the noise parameter precisely defined in terms of characteristic functions in Section~\ref{section-channels}. For both channels the minimal noise ($\mu_{\rm QL} \equiv \frac{1}{2}|\kappa - 1|$) corresponds to the quantum limited operation, the extra noise is $a = \mu - \mu_{\rm QL}$. The fact that the noise parameter $\mu > 0$ is a consequence of the canonical commutation relation for the photon creation and annihilation operators which is to be respected by the deterministic attenuation and amplification. The channel $\Phi(1,0)$ is the ideal channel (identity transformation), whereas $\Phi(1,\mu)$ describes an addition of classical noise. All the described channels are Gaussian and transform any Gaussian state into another Gaussian state. So does the tensor product $\Phi^A(\kappa_1,\mu_1) \otimes \Phi^B(\kappa_2,\mu_2)$ of Gaussian channels $\Phi(\kappa_1,\mu_1)$ and $\Phi(\kappa_2,\mu_2)$. The one-sided noise $\Phi^A(1,0) \otimes \Phi^B(\kappa_2,\mu_2)$ destroys any entanglement between modes $A$ and $B$ (i.e., the channel $\Phi(\kappa_2,\mu_2)$ is entanglement breaking) if $\mu_2 \geq \frac{1}{2}(\kappa_2 + 1)$. The region of parameters $\kappa_1$, $\mu_1$, $\kappa_2$, $\mu_2$, where the two-sided noise $\Phi^A(\kappa_1,\mu_1) \otimes \Phi^B(\kappa_2,\mu_2)$ annihilates entanglement is not fully characterized yet, and this is a goal of this paper to advance toward better understanding of this general case.

Given the fact that Gaussian states, Gaussian channels, and Gaussian measurements cover many practically relevant scenarios, it is not surprising the subfield of Gaussian quantum information emerged~\cite{weedbrook-2012}. The subfield is often thought of to be closed in the sense that an optimization problem for a Gaussian channel or a Gaussian measurement should have a solution within the class of Gaussian states or the Gaussian ensembles of Gaussian states. This is true, e.g., with regard to the maximal reliable communication rate through a quantum channel $\Phi(\kappa,\mu)$ (for all $\kappa$)~\cite{ghg-2015} as well as the maximal communication rate by using a Gaussian homodyne or heterodyne measurement~\cite{holevo-2021,holevo-2022,holevo-filippov-2022}. Therefore, it is very tempting to conjecture that the most robust entanglement with respect to Gaussian noises is exhibited by a Gaussian state~\cite{allegra-2010} (with the experiments being focused on the Gaussian states too~\cite{barbosa-2010,buono-2012}). The claim of Ref.~\cite{adesso-2011} is that if one adopts a bona fide measure of entanglement, which is continuous and strongly superadditive, such as the distillable entanglement; then at time $t$ in the Markovian Gaussian evolution $e^{Lt}$ (with generator $L$), the bipartite Gaussian state $e^{Lt}[\varrho_{12}^G(0)]$ with energy $\bar{n}(t)$ is the most entangled among all possible continuous-variable states $e^{Lt}[\varrho_{12}(0)]$ with the same energy, as a simple consequence of the maximum entropy property of Gaussian states~\cite{wolf-2006,van-enk-2005}. We believe this claim of Ref.~\cite{adesso-2011} is incorrect for the following reason. Ref.~\cite{wolf-2006} correctly mentions that if only the covariance matrix of some density operator $\varrho_1$ is known, then there exists a Gaussian density operator $\varrho_1^G$ with the same covariance matrix (hence, the same energy) such that the entropy of $\varrho_1^G$ is greater than the entropy of $\varrho_1$. If $\varrho_1$ and $\varrho_1^G$ are the reduced density operators of pure bipartite states $\varrho_{12}$ and $\varrho_{12}^G$, then the state $\varrho_{12}$ is indeed less entangled than the state $\varrho_{12}^G$~\cite{van-enk-2005}. However, this line of reasoning cannot be generalized to the evolution states $\varrho_{12}^G(t)$ and $\varrho_{12}(t)$ without breaking the equalities $\varrho_{12}^G(t) = e^{Lt}[\varrho_{12}^G(0)]$ and $\varrho_{12}(t) = e^{Lt}[\varrho_{12}(0)]$. In other words, the maximum entropy consideration does not guarantee that the states $\varrho_{12}^G(t)$ and $\varrho_{12}(t)$ are obtained from some legitimate states $\varrho_{12}^G(0)$ and $\varrho_{12}(0)$ in the same dynamical evolution $e^{Lt}$. In fact, Proposition~1 in Ref.~\cite{wolf-2006} is in accordance with our findings (that we present in this paper) as it states that a continuous and strongly superadditive entanglement measure $E$ satisfies the inequality $E(\varrho_{12}^G) \leq E(\varrho_{12})$ for every density operator $\varrho_{12}$ with finite covariance matrix and \emph{any} Gaussian state $\varrho_{12}^G$ with the same covariance matrix, i.e., Gaussian states give a lower bound for the entanglement $E(\varrho_{12})$. Finally, the conjecture on the superior robustness of Gaussian entanglement was shown to be false in 2011 by Sabapathy et al~\cite{sabapathy-2011}. They showed that none of two-mode Gaussian states can preserve entanglement under the Gaussian transformation $\Phi(\kappa,\mu) \otimes \Phi(\kappa,\mu)$ if $\mu \geq \frac{1}{2}$, whereas there exits a non-Gaussian state $\varrho_{12}(0)$ that remains entangled if $\kappa > \kappa_{\ast} \approx \frac{1}{2}$ and $\mu = \frac{1}{2}$. Moreover, for such parameters the preserved entanglement of the output state $\varrho_{12}(t)$ is \emph{distillable} [i.e., $E\big(\varrho_{12}(t)\big) > 0$ whereas $E\big(\varrho_{12}^G(t)\big) = 0$ for any initial Gaussian state $\varrho_{12}^G(0)$ with an arbitrary energy] because for the entanglement detection Sabapathy et al consider an effective two-qubit subspace (spanned by two-mode Fock states $\ket{00}$, $\ket{0n}$, $\ket{n0}$, $\ket{nn}$) and any non-zero two-qubit entanglement is known to be distillable~\cite{horodecki-1997}.

The result of Sabapathy et al~\cite{sabapathy-2011} was further strengthened in Ref.~\cite{filippov-ziman-2014}, where it was shown that the one-photon non-Gausian state $\ket{\psi_{\ast}^{AB}} = \frac{1}{\sqrt{2}} (\ket{01}-\ket{10})$ outperforms all Gaussian states (with an arbitrarily high energy) and remains entangled when affected upon by a symmetric noise $\Phi(\kappa,\mu) \otimes \Phi(\kappa,\mu)$ with $\mu < \frac{1}{2} \sqrt{1 + \kappa^2}$ for all $\kappa$. In the case of asymmetric local Gaussian noise $\Phi^A(\kappa_1,\mu_1) \otimes \Phi^B(\kappa_2,\mu_2)$, the question has remained open so far because the state $\ket{\psi_{\ast}^{AB}}$ cannot generally outperform any Gaussian state in terms of entanglement robustness. It is not hard to see with the help of Simon's criterion~\cite{simon-2000} that the Gaussian channel $\Phi^A(\kappa_1,\mu_1) \otimes \Phi^B(\kappa_2,\mu_2)$ may preserve a two-mode Gaussian entanglement (i.e., entanglement of some two-mode Gaussian state) if and only if~\cite{filippov-ziman-2014}
\begin{equation}\label{gaussian-inequality}
\kappa_1 \mu_2 + \kappa_2 \mu_1 < \frac{1}{2}(\kappa_1 + \kappa_2).
\end{equation}
\noindent The results of Ref.~\cite{filippov-ziman-2014} indicate that the state $\Phi^A(\kappa_1,\mu_1) \otimes \Phi^B(\kappa_2,\mu_2) [\ket{\psi_{\ast}^{AB}}\bra{\psi_{\ast}^{AB}}]$ can hardly remain entangled for all the parameters satisfying the inequality~\eqref{gaussian-inequality}, thereby leaving an open problem on whether Gaussian states may potentially be optimal for entanglement distribution through asymmetric local Gaussian channels. Here we finally resolve this question in negative by demonstrating a channel-dependent non-Gaussian state $\ket{\widetilde{\psi}_{c_{\kappa_1,\mu_1,\kappa_2,\mu_2}}^{AB}}$ that outperforms any Gaussian state and remains entangled under the transformation $\Phi^A(\kappa_1,\mu_1) \otimes \Phi^B(\kappa_2,\mu_2)$ whenever
\begin{equation}\label{main-inequality}
\kappa_1 \mu_2^2 + \kappa_2 \mu_1^2 < \frac{1}{4}(\kappa_1 + \kappa_2) (1 + \kappa_1 \kappa_2),
\end{equation}

\noindent which is a strictly larger region than \eqref{gaussian-inequality}. This result explicitly shows that the non-Gaussian entanglement is indeed more resilient to local Gaussian noises as compared to Gaussian states. Non-Gaussian entanglement can be useful not only in the entanglement distribution but also in other tasks, for example, the multi-component cat states were shown to outperform two-mode squeezed vacuum states in phase estimation~\cite{lee-2020}. 

The paper is organised as follows. In Section~\ref{section-channels}, we briefly review the formalism of characteristic functions and describe noisy attenuator and amplifier channels. In Section~\ref{section-ent-witness}, the entanglement witness used is presented. Section~\ref{section-robust} justifies the main result of the paper, the inequality \eqref{main-inequality}, and presents a family of channel-dependent non-Gaussian states $\ket{\widetilde{\psi}_{c_{\kappa_1,\mu_1,\kappa_2,\mu_2}}^{AB}}$. Summary of the results is given in Section~\ref{section-conclusions}.

\section{Gaussian attenuation and amplification channels} \label{section-channels}

A continuous-variable quantum state can be alternatively described by a density operator $\varrho$ (positive semidefinite operator with unit trace) or any of numerous phase-space functions, for instance, the characteristic function $\varphi({\bf z}) = {\rm tr}[\varrho W({\bf z})]$, where the Weyl operator $W({\bf z})$ for $N$ modes is expressed through the canonical operators $q_i$ and $p_i$ (satisfying the canonical commutation relation $[q_i,p_j] = i \delta_{ij}$) through
\begin{equation}
W({\bf z}) = \exp[i(q_1 x_1 + p_1 y_1 + \ldots + q_N x_n + p_N y_N)],
\end{equation}

\noindent where $(x_1,y_1,\ldots,x_N,y_N)^{\top} \equiv {\bf z}$ are coordinates in the real symplectic space $(\mathbb{R}^{2N},\Delta)$ with the symplectic form
\begin{equation*}
\Delta = \bigoplus\limits_{i=1}^N \left( \begin{array}{cc}
                                                              0 & -1 \\
                                                              1 & 0
                                                            \end{array}
\right).
\end{equation*}
A Gaussian state has a Gaussian characteristic function $\varphi({\bf z}) = \exp( i {\bf l}^{\top} {\bf z} - \frac{1}{2} {\bf z}^{\top} {\bf V} {\bf z})$, where ${\bf l} = (\braket{q_1},\braket{p_1},\ldots,\braket{q_N},\braket{p_N})^{\top}$ is a $2N$-dimensional column of the first moments and ${\bf V}$ is the $2N \times 2N$ covariance matrix with elements $V_{kl} = \frac{1}{2}(\braket{R_k R_l} - \braket{R_k}\braket{R_l})$, $R_k,R_l \in (q_1,p_1,\ldots,q_N,p_N)$. This implies that the Gaussian state is fully determined by the first and second moments of the canonical operators.

The Gaussian channel is a completely positive and trace preserving map that maps any Gaussian state into a Gaussian one and acts as follows in terms of input and output characteristic functions $\varphi_{\rm in}({\bf z})$ and $\varphi_{\rm out}({\bf z})$~\cite{holevo-2019}:
\begin{equation}\label{Gaussian-channel}
\varphi_{\rm out}({\bf z}) = \varphi_{\rm in}({\bf K}{\bf z}) \exp\left(i {\bf m}^{\top} {\bf z} - \frac{1}{2} {\bf z}^{\top} {\bf M} {\bf z} \right),
\end{equation}

\noindent where the real scaling matrix ${\bf K}$ and the real noise-quantifying symmetric matrix ${\bf M}$ must satisfy the inequality ${\bf M} \geq \frac{i}{2} (\Delta - {\bf K}^{\top} \Delta {\bf K})$ for the map to be completely positive. The full characterization of one-mode Gaussian channels is given in Ref.~\cite{holevo-2007}. We focus on the most important (from the viewpoint of physical applications) deterministic Gaussian channels: phase-insensitive attenuators and amplifiers. These two types of Gaussian channels can be both described within unified formulas: ${\bf K} = \sqrt{\kappa}\left(
                                                                      \begin{array}{cc}
                                                                        1 & 0 \\
                                                                        0 & 1 \\
                                                                      \end{array}
                                                                    \right)
$, ${\bf m} = 0$, ${\bf M} = \mu \left(
                                                                      \begin{array}{cc}
                                                                        1 & 0 \\
                                                                        0 & 1 \\
                                                                      \end{array}
                                                                    \right)
$, where $0 \leq \kappa < 1$ for the attenuator, $\kappa > 1$ for the amplifier, and the noise parameter $\mu \geq \frac{1}{2}|1 - \kappa| \equiv \mu_{\rm QL}$ in both cases.  We will denote such one-mode Gaussian channels by $\Phi(\kappa,\mu)$. The minimal noise $\mu_{\rm QL}$ corresponds to the quantum-limited operation in which the admixed noise originates from vacuum fluctuations of the environment~\cite{caves-1982}; however, in some physically relevant systems the noise $\mu$ can be much higher that $\mu_{\rm QL}$ due to a high-temperature environment~\cite{isar-2009}, for instance, in microwave quantum experiments~\cite{mallet-2011,eichler-2011,filippov-manko-2011}.

\begin{figure}
\centering
\includegraphics[width=7cm]{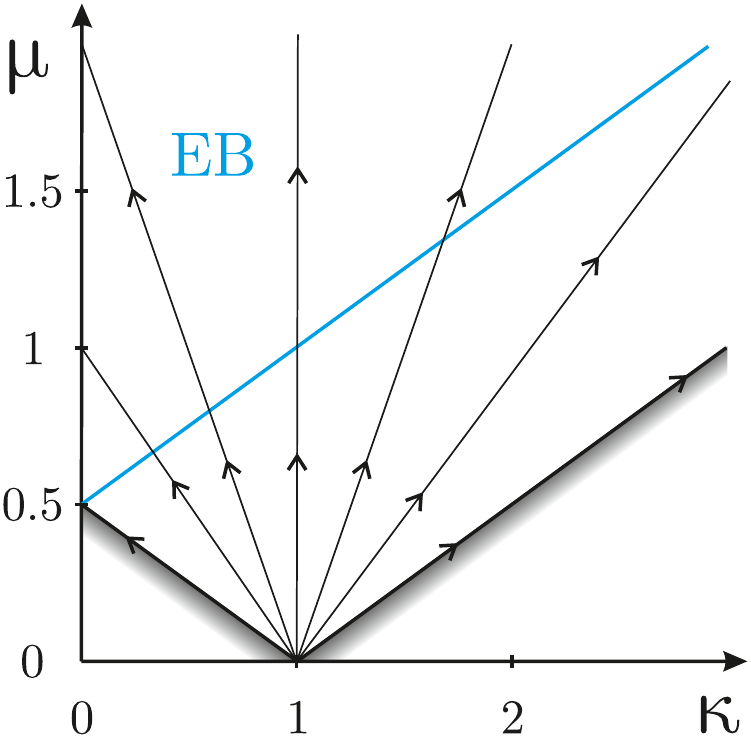}
\caption{\centering The power attenuation/amplification factor $\kappa$ and the noise parameter $\mu$ for a legitimate one-mode quantum channel $\Phi(\kappa,\mu)$. $\Phi(1,0)$ is the identity transformation. The channel is entanglement breaking (so that the one-sided noise $\Phi(1,0) \otimes \Phi(\kappa,\mu)$ destroys any two-mode entanglement) above the blue line. Arrows indicate how the parameters $\kappa$ and $\mu$ change in time in the semigroup dynamics.} \label{figure1}
\end{figure}

The channel $\Phi(\kappa,\mu)$ is known to be entanglement-breaking (so that the one-sided noise $\Phi(1,0) \otimes \Phi(\kappa,\mu)$ destroys any two-mode entanglement) if and only if the noise parameter $\mu$ exceed the threshold value, $\mu \geq \frac{1}{2}(\kappa + 1)$~\cite{holevo-2008}. This value is depicted in Fig.~\ref{figure1}.

Clearly, the quantum channel $\Phi(\kappa,\mu)$ with fixed parameters $\kappa$ and $\mu$ represents a snapshot of the dynamical map at a particular time moment $t$, which may correspond to a finite propagation time through a communication line. In a true time evolution the parameters $\kappa$ and $\mu$ become functions of time $t$, $\kappa(t)$ and $\mu(t)$. For instance, in the semigroup attenuation or amplification dynamics $\Phi = e^{Lt}$ with the generator $L$ \cite[section 3.4.6]{breuer-2002} we obtain the following dependencies:
\begin{equation}\label{GKSL}
\kappa(t) = e^{\pm\Gamma t}, \qquad \mu(t) = \pm \left(e^{\pm \Gamma t} - 1 \right) \left( \bar{n} + \frac{1}{2} \right),
\end{equation}

\noindent where the sign $+$ ($-$) describes amplification (attenuation), $\Gamma \geq 0$ is the process rate, and $\bar{n}$ is the average number of thermal photons in the environment. For such a semigroup dynamics the one-parameter family of maps $\Phi\big( \kappa(t),\mu(t) \big)$ is associated with a straight line in the parameter space $(\kappa,\mu)$, see Fig.~\ref{figure1}.

The diagonal sum representation for the channel $\Phi(\kappa,\mu)$ is found in Ref.~\cite{ivan-2011} in terms of the Fock states $\{\ket{n}\}_{n=0}^{\infty}$ and in Ref.~\cite{filippov-ziman-2014} in terms of the coherent states $\{\ket{\alpha}\}_{\alpha \in \mathbb{C}}$. We use the latter one as it sheds more light on the phase space picture~\cite{schleich,manko-2005}. The integral form of the diagonal sum representation reads
\begin{equation}\label{Phi-through-K}
\Phi(\kappa,\mu) [\varrho] = \frac{1}{\pi^2} \iint A_{\alpha\beta}(\kappa,\mu) \varrho A_{\alpha\beta}^{\dag}(\kappa,\mu) d^2\alpha d^2\beta,
\end{equation}

\noindent where the Kraus operators $A_{\alpha\beta}(\kappa,\mu)$ rather nontrivially depend on parameters $\kappa$ and $\mu$ through auxiliary expressions
\begin{eqnarray}
  \tau \!\!\! &=& \!\!\! \max(1,\kappa) + \mu - \frac{1}{2}|\kappa - 1|, \qquad \eta = \frac{\kappa}{\tau}, \label{tau-eta}\\
  A_{\alpha\beta}(\kappa,\mu) \!\!\! &=& \!\!\!
 \int \frac{d^2 \gamma}{\pi\sqrt{\tau}} \exp \bigg[ -\frac{|\alpha|^2+|\beta|^2+|\gamma|^2}{2} + \sqrt{1-\eta} \, \alpha\gamma + \frac{1}{2\tau} \left|
\sqrt{\tau - 1} \, \beta + \sqrt{\eta} \, \gamma
\right|^2  \bigg] \nonumber\\
&& \times \bigg| \sqrt{\frac{\tau - 1}{\tau}} \,
\beta+\sqrt{\frac{\eta}{\tau}} \, \gamma \bigg\rangle
\bra{\gamma}.
\end{eqnarray}

\noindent The peculiar form of the presented Kraus operators originates from the concatenation formula $\Phi(\kappa,\mu) = \Phi_{{\rm QL} \tau}
\circ \Phi_{{\rm QL}\eta}$ that shows the channel $\Phi(\kappa,\mu)$ can always be thought of as a sequential application of the quantum limited attenuator $\Phi_{{\rm QL} \eta}$ (with the attenuation factor $\eta$) and the quantum limited amplifier $\Phi_{{\rm QL} \tau}$ (with the power gain $\tau$). Using the diagonal sum representation above it is not hard to see the effect of the channel $\Phi(\kappa,\mu)$ on the outer product of coherent states, namely,
\begin{eqnarray}
&& \Phi(\kappa,\mu)[\ket{\gamma}\bra{\delta}] = \int \frac{d^2 \sigma}{\pi \tau} \, f_{\gamma\delta}(\sigma) \bigg| \sqrt{\frac{\eta}{\tau}} \, \gamma + \sqrt{\frac{\tau - 1}{\tau}} \, \sigma \bigg\rangle  \bigg\langle \sqrt{\frac{\eta}{\tau}} \, \delta + \sqrt{\frac{\tau - 1}{\tau}} \, \sigma \bigg|, \label{Phi-on-op}\\
&& f_{\gamma\delta}(\sigma) = \exp \left[ \left(\frac{\eta}{\tau} - 1 \right) \frac{|\gamma|^2+|\delta|^2}{2} + (1-\eta)\gamma\delta^{\ast} + \frac{\sqrt{\eta(\tau - 1)}}{\tau} {\rm Re}[\sigma^{\ast}(\gamma+\delta)] - \frac{|\sigma|^2}{\tau} \right]. \label{f-op} \nonumber\\
\end{eqnarray}

\noindent Since any density operator $\varrho$ can be decomposed as $\varrho = \pi^{-2} \iint d^2 \gamma d^2\delta \bra{\gamma}\varrho\ket{\delta} \ket{\gamma}\bra{\delta}$, equations \eqref{Phi-on-op} and \eqref{f-op} define the effect of the Gaussian channel $\Phi(\kappa,\mu)$ on any state (including non-Gaussian ones).

\section{Entanglement witness} \label{section-ent-witness}

A bipartite state $\varrho^{AB}$ is called entangled if it cannot be represented by a convex sum of density operators $\varrho_k^A$ and $\varrho_k^B$ for individual subsystems $A$ and $B$. Otherwise the state is separable. In our case each subsystem is a mode of electromagnetic radiation, so we analyze a two-mode entanglement. Although simple and physically clear entanglement criterions are known for Gaussian two-mode systems~\cite{simon-2000,dodonov-2005}, they are still missing for non-Gaussian states. For this reason we resort to an entanglement witness formalism that enables us to detect entanglement within non-Gaussian states too~\cite{filippov-ziman-2014}.

Consider a Hermitian operator
\begin{equation}
\label{witness} W_{\lambda} = \frac{1}{\pi^2} \iint {d^2\alpha} \,
{d^2\beta} \, e^{\lambda(|\alpha|^2+|\beta|^2)}
\ket{\alpha}\bra{\beta} \otimes \ket{\beta}\bra{\alpha},
\end{equation}

\noindent which reduces to a swapping operator if $\lambda = 0$. The important feature of this operator with $\lambda \in \mathbb{R}$ is that its average value is nonnegative in any separable state because
\begin{equation}\label{W-justification}
\bra{\phi^A} \otimes \bra{\chi^B} W_{\lambda} \ket{\phi^A} \otimes \ket{\chi^B} = \left| \frac{1}{\pi} \iint {d^2\alpha} \, e^{\lambda |\alpha|^2}
\braket{\phi^A | \alpha}\braket{\alpha | \chi^B} \right|^2 \geq 0
\end{equation}

\noindent for all vectors $\ket{\phi^A}$ and $\ket{\chi^B}$. Therefore, $W_{\lambda}$ is an entanglement witness in the sense that the negativity in the average value ${\rm tr}[\varrho^{AB} W_{\lambda}]$ can only arise from the entangled state $\varrho^{AB}$, so ${\rm tr}[\varrho^{AB} W_{\lambda}] < 0$ unambiguously indicates the entanglement between subsystems $A$ and $B$. If $\lambda > 0$, then $W_{\lambda}$ is an unbounded operator; however, the average value ${\rm tr}[\varrho^{AB} W_{\lambda}]$ may still be finite.

Suppose the local Gaussian channel $\Phi^A(\kappa_1,\mu_1) \otimes \Phi^B(\kappa_2,\mu_2)$ acts on some two-mode pure state $\ket{\psi^{AB}}\bra{\psi^{AB}}$, then the output state $\varrho_{AB} = \Phi^A(\kappa_1,\mu_1) \otimes \Phi^B(\kappa_2,\mu_2)[\ket{\psi^{AB}}\bra{\psi^{AB}}]$ is verified to be entangled if ${\rm tr}[\varrho^{AB} W_{\lambda}]< 0$ for some $\lambda \in \mathbb{R}$. One can actually adjust the input state $\ket{\psi^{AB}}\bra{\psi^{AB}}$ to be dependent on the channel parameters $\kappa_1$, $\mu_1$, $\kappa_2$, $\mu_2$ in order to find a robust entangled state for a given channel.

\section{Robust non-Gaussian entanglement} \label{section-robust}

Here we present a non-Gaussian state $\ket{\psi_c^{AB}}\bra{\psi_c^{AB}}$, which exhibits high resilience to the local Gaussian noise $\Phi^A(\kappa_1,\mu_1) \otimes \Phi^B(\kappa_2,\mu_2)$. A real scaling coefficient $c \in \mathbb{R}$ describes asymmetry in the skewed two-mode ``cat state''
\begin{equation}\label{psi-resilient}
\ket{\psi_c^{AB}}  = \frac{\ket{\gamma}\otimes\ket{0} -  \ket{0} \otimes \ket{c\gamma}}{\sqrt{2 \left[1 - \exp\left(- (1+c^2) \frac{|\gamma|^2}{2} \right)\right]}},
\end{equation}

\noindent where $\ket{\gamma}$ and $\ket{c \gamma}$ are coherent states and $\ket{0}$ is a vacuum state. At the end of this section we will make the coefficient $c$ dependent on parameters $\kappa_1$, $\mu_1$, $\kappa_2$, $\mu_2$ so as to present the noise-specific resilient entangled state; however, for now we treat it as an independent parameter. If $c = 1$, then we get a symmetric state whose entanglement is robuster against symmetric Gaussian noise $\Phi^A(\kappa,\mu) \otimes \Phi^B(\kappa,\mu)$ than that of any Gaussian state~\cite{filippov-ziman-2014}. When we deal with asymmetric noise $\Phi^A(\kappa_1,\mu_1) \otimes \Phi^B(\kappa_2,\mu_2)$, the robustest input state is expected to be asymmetric too and this is the physical reason to consider the skewed state \eqref{psi-resilient}.

The output state $\Phi^A(\kappa_1,\mu_1) \otimes \Phi^B(\kappa_2,\mu_2)[\ket{\psi_c^{AB}}\bra{\psi_c^{AB}}]$ is a linear combination of four operators
\begin{eqnarray}
&& \Phi(\kappa_1,\mu_1)[\ket{\gamma}\bra{\gamma}] \otimes \Phi(\kappa_2,\mu_2)[\ket{0}\bra{0}], \label{out-1} \\
&& \Phi(\kappa_1,\mu_1)[\ket{\gamma}\bra{0}] \otimes \Phi(\kappa_2,\mu_2)[\ket{0}\bra{c \gamma}], \\
&& \Phi(\kappa_1,\mu_1)[\ket{0}\bra{\gamma}] \otimes \Phi(\kappa_2,\mu_2)[\ket{c\gamma}\bra{0}], \\
&& \Phi(\kappa_1,\mu_1)[\ket{0}\bra{0}] \otimes \Phi(\kappa_2,\mu_2)[\ket{c \gamma}\bra{c \gamma}]. \label{out-4}
\end{eqnarray}

\noindent Each of the operators \eqref{out-1}--\eqref{out-4} is readily calculated via Eq.~\eqref{Phi-on-op} and then used to calculate the Hilbert-Schmidt scalar product with the entanglement witness $W_{\lambda}$:
\begin{eqnarray}
&& {\rm tr} \left\{ W_{\lambda} \, \Phi(\kappa_1,\mu_1)[\ket{\gamma}\bra{\gamma}] \otimes \Phi(\kappa_2,\mu_2)[\ket{0}\bra{0}] \right\}, \label{out-W-1} \\
&& {\rm tr} \left\{ W_{\lambda} \Phi(\kappa_1,\mu_1)[\ket{\gamma}\bra{0}] \otimes \Phi(\kappa_2,\mu_2)[\ket{0}\bra{c \gamma}]  \right\}, \\
&& {\rm tr} \left\{ W_{\lambda} \Phi(\kappa_1,\mu_1)[\ket{0}\bra{\gamma}] \otimes \Phi(\kappa_2,\mu_2)[\ket{c \gamma}\bra{0}] \right\}, \\
&& {\rm tr} \left\{ W_{\lambda} \Phi(\kappa_1,\mu_1)[\ket{0}\bra{0}] \otimes \Phi(\kappa_2,\mu_2)[\ket{c \gamma}\bra{c \gamma}]  \right\}. \label{out-W-4}
\end{eqnarray}

\noindent The straightforward integral calculations show that all four expressions are finite if
\begin{equation}\label{lambda-restricion}
\lambda < 1 - \sqrt{\frac{(\tau_1 -1 )(\tau_2 - 1)}{\tau_1 \tau_2}}.
\end{equation}

\noindent Suppose $\gamma \rightarrow 0$, then it is not hard to see via the Taylor series that the condition
\begin{equation}\label{ineq-main}
{\rm tr} \left\{ W_{\lambda} \, \Phi^A(\kappa_1,\mu_1) \otimes \Phi^B(\kappa_2,\mu_2)[\ket{\psi_c^{AB}}\bra{\psi_c^{AB}}] \right\} < 0
\end{equation}

\noindent is fulfilled if the following inequality is met for the leading term (proportional to $|\gamma|^2$):
\begin{eqnarray}
&&  b_1 \lambda^2 -2 b_2 \lambda + b_3 < 0, \label{ineq-lambda-negative-ew} \\
&&  b_1 = 1 - \eta_1 + c^2 (1 - \eta_2), \\
&&  b_2 = b_1 - c \sqrt{\frac{\eta_1\eta_2}{\tau_1\tau_2}}, \\
&&  b_3 = \frac{[\tau_1(1 - \eta_1) + \tau_2 - 1] - 2 c \sqrt{\eta_1 \tau_1 \eta_2 \tau_2} + c^2 [\tau_1 + \tau_2 (1-\eta_2) -1]}{\tau_1 \tau_2}, \quad
\end{eqnarray}

\noindent where $\tau_i,\eta_i$ are expressed through $\kappa_i,\mu_i$ via equation \eqref{tau-eta}, $i= 1,2$. Some elementary algebra yields that the inequality \eqref{ineq-lambda-negative-ew} is fulfilled for some $\lambda$ satisfying \eqref{lambda-restricion} if
\begin{equation}\label{tau-condition}
\tau_1 < \frac{1 + c^2(1-\eta_2)}{1-\eta_1+c^2(1-\eta_2)} \text{~~~and~~~} \tau_2 < \frac{1 - \eta_1 + c^2}{1-\eta_1+c^2(1-\eta_2)}.
\end{equation}

Since the limit $\gamma \rightarrow 0$ has been used in the derivation above, we actually get the one-photon continuous-variable state
\begin{equation}\label{psi-resilient-1-photon}
\ket{\widetilde{\psi}_c^{AB}}  = \frac{\ket{1}\otimes\ket{0} -  c \ket{0} \otimes \ket{1}}{\sqrt{1+c^2}},
\end{equation}

\noindent where $\ket{1}$ is a single-photon Fock state. Inequalities \eqref{tau-condition} give sufficient conditions under which the output state $\Phi^A(\kappa_1,\mu_1) \otimes \Phi^B(\kappa_2,\mu_2)[\ket{\widetilde{\psi}_c^{AB}}\bra{\widetilde{\psi}_c^{AB}}]$ is entangled. These inequalities can be rewritten in terms of the additional noise on top of the quantum-limited operation, $a_i = \mu_i - \frac{1}{2}|\kappa_i - 1| \geq 0$, $i = 1,2$, as the following result.

\begin{proposition} \label{proposition-c-noises}
The output state $\Phi^A(\kappa_1,\mu_1) \otimes \Phi^B(\kappa_2,\mu_2)[\ket{\widetilde{\psi}_c^{AB}}\bra{\widetilde{\psi}_c^{AB}}]$ is entangled if
\begin{itemize}
\item both $\Phi^A(\kappa_1,\mu_1)$ and $\Phi^B(\kappa_2,\mu_2)$ are attenuators ($\kappa_{1,2} \leq 1$) and
\begin{equation}\label{attenuator-attenuator-condition-on-a}
a_1 < \frac{\kappa_1(1+a_2)}{(c^2 + 1)(1+a_2) - c^2 \kappa_2}, \qquad a_2 < \frac{c^2 \kappa_2 (1+a_1)}{(c^2+1)(1+a_1)-\kappa_1};
\end{equation}
\item $\Phi^A(\kappa_1,\mu_1)$ is an attenuator ($\kappa_{1} \leq 1$) and $\Phi^B(\kappa_2,\mu_2)$ is an amplifier ($\kappa_{2} \geq 1$) and
\begin{equation}\label{attenuator-amplifier-condition-on-a}
a_1 < \frac{\kappa_1(\kappa_2+a_2)}{\kappa_2 + (c^2 + 1)a_2}, \qquad a_2 < 1 - \frac{\kappa_2 (1+a_1 - \kappa_1)}{(c^2+1)(1+a_1)-\kappa_1};
\end{equation}
\item both $\Phi^A(\kappa_1,\mu_1)$ and $\Phi^B(\kappa_2,\mu_2)$ are amplifiers ($\kappa_{1,2} \geq 1$) and
\begin{equation}\label{attenuator-attenuator-condition-on-a}
a_1 < 1 - \frac{c^2 \kappa_1 a_2}{\kappa_2 + (c^2 + 1)a_2}, \qquad a_2 < 1 - \frac{\kappa_2 a_1}{c^2 \kappa_1 + (c^2+1)a_1},
\end{equation}
\end{itemize}
\noindent where $a_i = \mu_i - \frac{1}{2}|\kappa_i - 1| \geq 0$, $i = 1,2$.
\end{proposition}

\begin{figure}
\centering
\includegraphics[width=18cm]{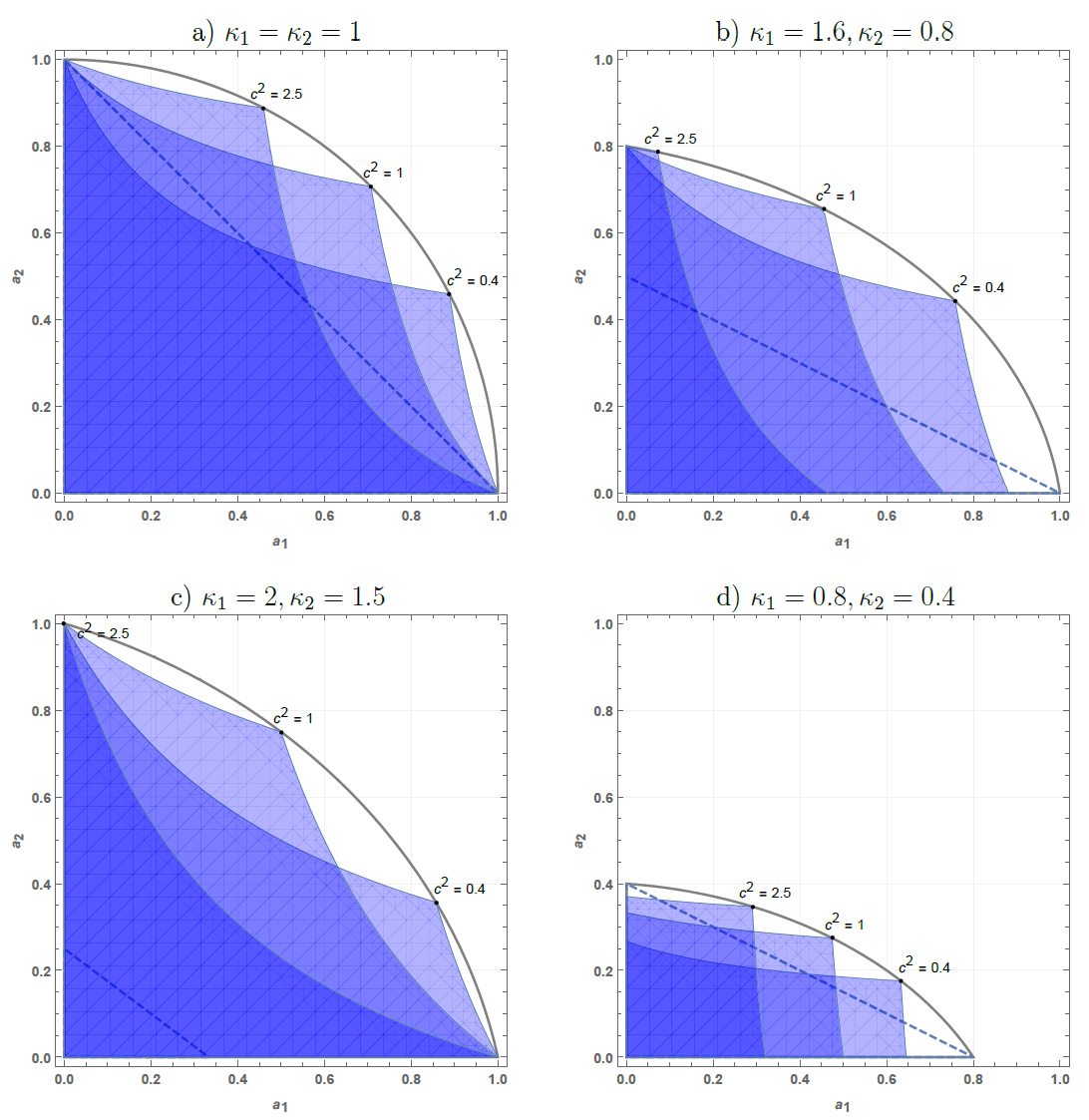}
\caption{Region of the extra noise $a_i$ in the Gaussian attenuator/amplifier with the attenuation factor/power gain $\kappa_i$, $i=1,2$, within which the non-Gaussian state $\ket{\widetilde{\psi}_c^{AB}}$ remains entangled. Dashed line corresponds to the boundary attainable with Gaussian states.}
\label{figure2}
\end{figure}

If $c=1$ and we deal with a symmetric state $\ket{\widetilde{\psi}_1} = \frac{1}{\sqrt{2}} \left( \ket{1}\ket{0} - \ket{0}\ket{1} \right)$, then Proposition~\ref{proposition-c-noises} reproduces the known results from Ref.~\cite{filippov-ziman-2014}; however, if $c \neq 1$, then Proposition~\ref{proposition-c-noises} gives a valuable generalization. In Fig.~\ref{figure2} we depict typical noise levels tolerated by the state $\ket{\widetilde{\psi}_c}$ without losing entanglement. If we choose specific values of $\kappa_1$, $\kappa_2$, and $c$, then the shaded region in Fig.~\ref{figure2} shows the extra noises $a_1$, $a_2$, which do not destroy entanglement of $\ket{\widetilde{\psi}_c}$. One can clearly see that the non-Gaussian state $\ket{\widetilde{\psi}_c}$ outperforms all Gaussian ones (dashed line) if we deal with a two-side (asymmetric or symmetric) amplification [Fig.~\ref{figure2}(c)]; however, a single state $\ket{\widetilde{\psi}_c}$ with some fixed value of $c$ cannot outperform all Gaussian states (dashed line) in the case of asymmetric attenuation [$\kappa_{1,2}< 1$, $\kappa_1 > \kappa_2$, Fig.~\ref{figure2}(d)] or in the case of amplification-attenuation channel [$\kappa_1 > 1 > \kappa_2$, Fig.~\ref{figure2}(b)]. If $a_2$ tends to zero, then $\Phi^B(\kappa_2,\mu_2)$ approaches the quantum limited attenuation and the high-energy Gaussian entanglement tolerates this noise. Dashed line in Fig.~\ref{figure2} corresponds to the infinite-energy two-mode squeezed state~\cite{filippov-ziman-2014}. Nonetheless, varying the parameter $c$, we cover the area in the extra noise space $(a_1,a_2)$ which is strictly larger than the area for the Gaussian states [dashed line, Eq.~\eqref{gaussian-inequality}] and fully comprises the latter. Some elementary algebra shows that the envelope curve for the former area (the arc of an ellipsoid in Fig.~\ref{figure2}) is exactly given by the equation $\kappa_1 \mu_2^2 + \kappa_2 \mu_1^2 = \frac{1}{4}(\kappa_1 + \kappa_2) (1 + \kappa_1 \kappa_2)$. Therefore, if the inequality~\eqref{main-inequality} is fulfilled, then there exists some $c = c_{\kappa_1 \mu_1 \kappa_2 \mu_2}$ such that the state $\Phi^A(\kappa_1,\mu_1) \otimes \Phi^B(\kappa_2,\mu_2)[\ket{\widetilde{\psi}_c^{AB}}\bra{\widetilde{\psi}_c^{AB}}$ is entangled.

\begin{figure}
\centering
\includegraphics[width=18cm]{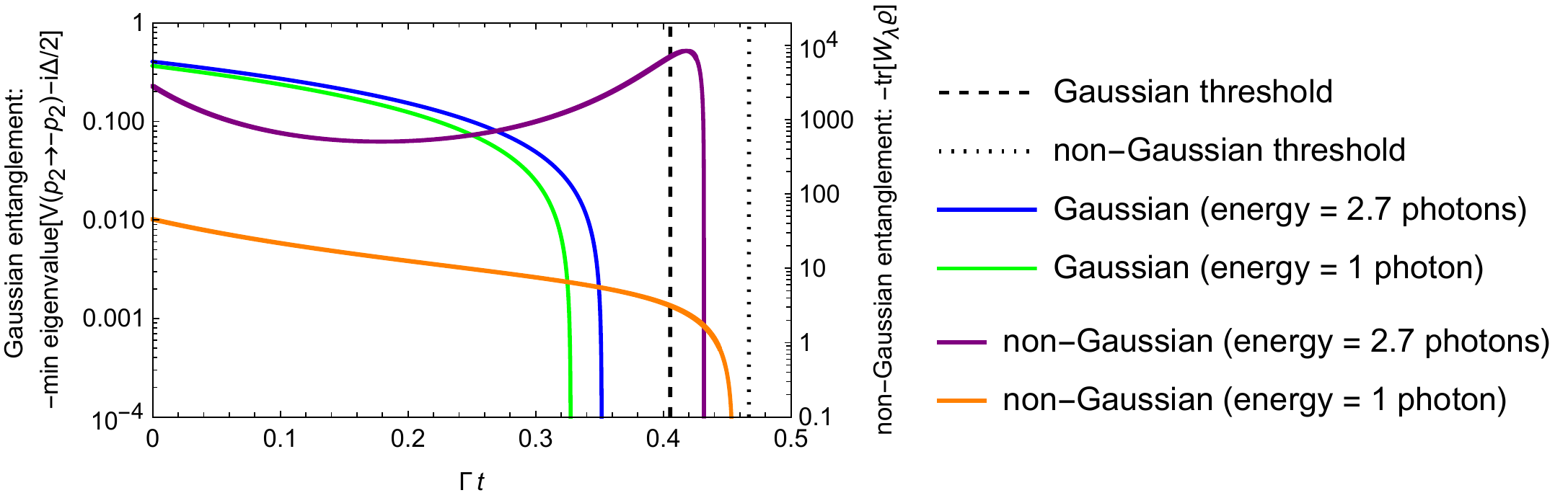}
\caption{Entanglement dynamics for Gaussian and non-Gaussian states in Example~\ref{example-1} vs dimensionless time. Inequalities \eqref{gaussian-inequality} and \eqref{main-inequality} establish thresholds for general Gaussian and specific non-Gaussian entanglement death time, respectively.}
\label{figure3}
\end{figure}

\begin{example} \label{example-1}
Consider the Gaussian local attenuation dynamics $e^{L_1 t} \otimes e^{L_2 t}$ with the time-dependent parameters \eqref{GKSL}, where the rates $\Gamma_1 = \Gamma_2 = \Gamma$ and $\bar{n}_1 = 0.5$, $\bar{n}_2 = 1.5$. Then
\begin{equation*}
\kappa_1(t) = \kappa_2(t) = e^{-\Gamma t}, \qquad \mu_1(t) = 1 - e^{-\Gamma t}, \qquad \mu_2(t) = 2(1 - e^{-\Gamma t}).
\end{equation*}
In such dynamics, the fundamental threshold \eqref{gaussian-inequality} implies that no Gaussian entanglement can survive if $\Gamma t \geq - \ln \frac{2}{3} \approx 0.405$. On the other hand, the inequality \eqref{main-inequality} implies that the non-Gaussian entanglement of states \eqref{psi-resilient} may be preserved up to the threshold $\Gamma t = - \ln \frac{10-\sqrt{19}}{9} \approx 0.467$. In Fig.~\ref{figure3} we show both thresholds as well as the dynamics of entanglement for Gaussian and non-Gaussian states. Non-Gaussian entanglement is characterized by the introduced entanglement witness, namely, the quantity $-{\rm tr}[W_{\lambda} \varrho(t)]$ with $\lambda = 0.74$. As initial states we use the non-Gaussian state \eqref{psi-resilient} with $\gamma = 1$ and $c = 2$, which has the energy $2.7$ photons, and the single-photon non-Gaussian state \eqref{psi-resilient-1-photon} with $c = 2$. Fig.~\ref{figure3} suggests that these non-Gaussian states remain entangled after passing the Gaussian threshold, with the single-photon state exhibiting stronger resilience for the entanglement measure used. The latter fact is in agreement with the earlier observations on the low-energy entanglement robustness~\cite{van-enk-2005}. (Non-monotonic behavior of our entanglement measure for the higher-energy state should not be surprising as the witness operator $W_\lambda$ assigns higher values to coherent states with higher amplitude. As the noise increases, the overlap with high-amplitude states increases too. In this sense, our entanglement measure is not an entanglement monotone but it is rather a very sensitive entanglement detector for slightly entangled states.) As examples of Gaussian states we consider two-mode squeezed vacuum states with the same initial energy ($2.7$ photons and $1$ photon). Unfortunately, our entanglement measure diverges for these states so we resort to the Simon criterion (necessary and sufficient condition for two-mode Gaussian entanglement~\cite{simon-2000}), namely, the entanglement measure is the minimal eigenvalue of the matrix ${\bf V}(p_2 \rightarrow - p_2) - \frac{i}{2} \Delta$, where ${\bf V}$ is the covariance matrix and ${\bf V}(p_2 \rightarrow - p_2)$ is its modified version obtained via inversion of the momentum operator for the second mode. Fig.~\ref{figure3} illustrates that the entanglement of such Gaussian states vanishes before the Gaussian threshold. The entanglement death time approaches the Gaussian threshold if the energy of the two-mode squeezed state tends to infinity. Whatever low the non-Gaussian entanglement can be after passing the Gaussian threshold, Ref.~\cite{sabapathy-2011} suggests that this entanglement is distillable in general, so it can be converted into a useful form. In principle, there may exist non-Gaussian states that remain entangled even after the established non-Gaussian threshold, because this threshold was derived for a specific family of non-Gaussian states $\{\ket{\psi_c^{AB}}\}$ and a specific entanglement measure $-{\rm tr}[W_{\lambda} \varrho(t)]$.
\end{example}

\begin{figure}
\centering
\includegraphics[width=8cm]{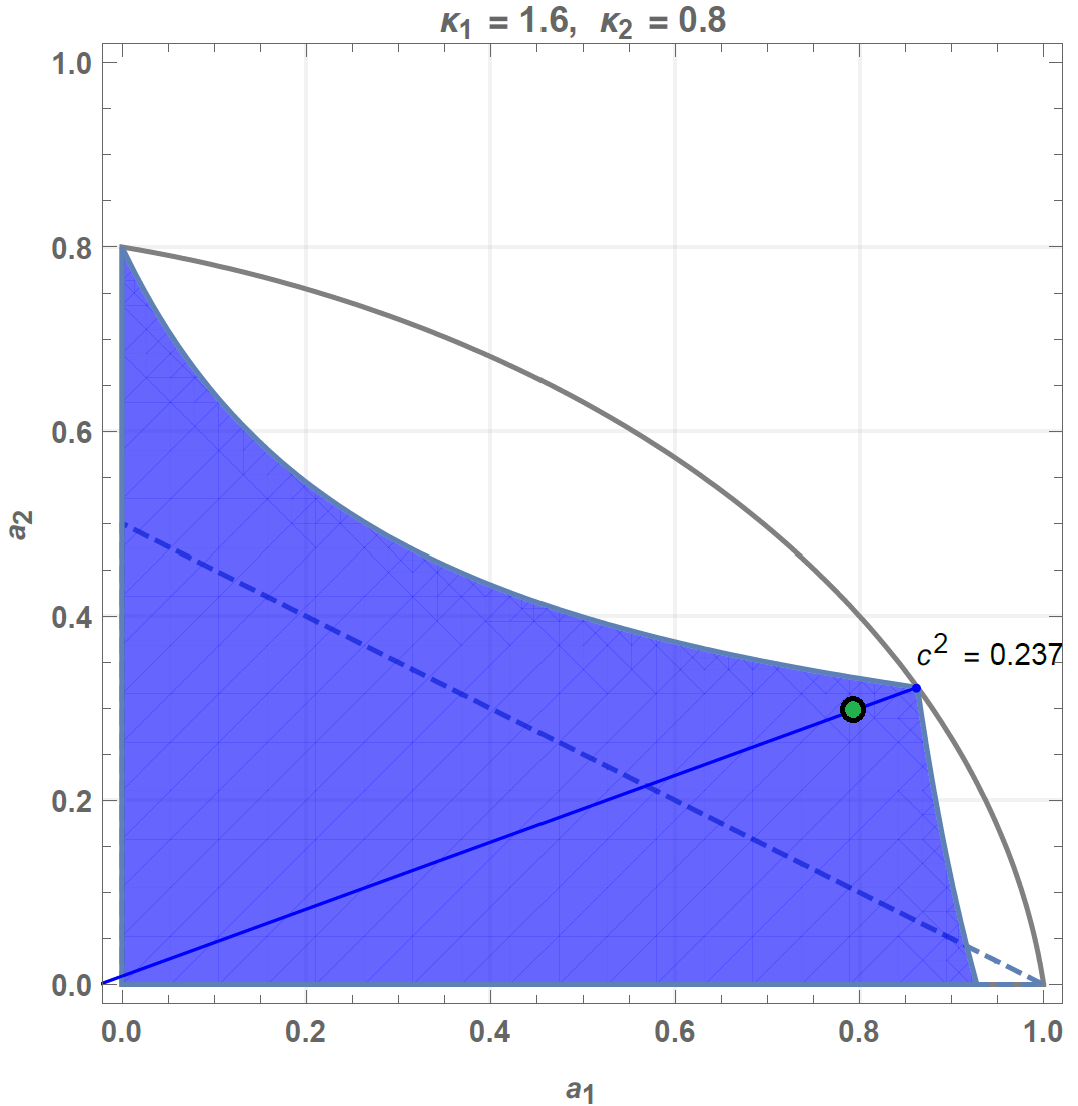}
\caption{Example on how to find the advantageous state $\ket{\widetilde{\psi}_{c_{\kappa_1 \mu_1 \kappa_2 \mu_2}}^{AB}}$, whose entanglement is resilient against a given extra noise $(a_1,a_2)$ (green dot) in the two-mode Gaussian attenuator/amplifier with the attenuation factor/power gain $\kappa_1,\kappa_2$ in the corresponding mode. Straight solid line passes through the ellipsoid center and the green dot, thus intersecting the ellipsoid \eqref{main-inequality} in the point $(\widetilde{a}_1,\widetilde{a}_2)$ that determines $c_{\kappa_1 \mu_1 \kappa_2 \mu_2}$ via formula~\eqref{c-optimal}.}
\label{figure4}
\end{figure}

To make the analysis complete and more applicable to practice we specify how the advantageous parameter $c_{\kappa_1 \mu_1 \kappa_2 \mu_2}$ should be chosen for given channel parameters $\kappa_1$, $\mu_1$, $\kappa_2$, $\mu_2$. Suppose the channel parameters $\kappa_1$, $\mu_1$, $\kappa_2$, $\mu_2$ satisfy the inequality \eqref{main-inequality}, then we fix the amplification/attenuation parameters $\kappa_{1,2}$ and find the maximally permissible noise levels $\widetilde{\mu}_{1,2}$ by projecting the point $(\mu_1,\mu_2)$ onto the boundary of the ellipsoid (see Fig.~\ref{figure4}), namely,
\begin{equation}\label{mu-projection}
\widetilde{\mu}_i = \mu_i \frac{\sqrt{\frac{1}{4}(\kappa_1 + \kappa_2)(1+ \kappa_1 \kappa_2)}}{\sqrt{\kappa_1 \mu_2^2 + \kappa_2 \mu_1^2}}, \qquad i =1,2.
\end{equation}

\noindent Then the advantageous parameter $c_{\kappa_1 \mu_1 \kappa_2 \mu_2}$ is chosen in such a way that a corner point for the noise-toleration figure coincides with the point $(\widetilde{a}_1,\widetilde{a}_2)$ in Fig.~\ref{figure4}, $\widetilde{a}_i = \widetilde{\mu}_i - \frac{1}{2}|\kappa_i - 1|$. The algebraic expression reads
\begin{equation}\label{c-optimal}
c_{\kappa_1 \mu_1 \kappa_2 \mu_2} = \sqrt{\frac{\widetilde{\tau}_2 - \widetilde{\tau}_1 + \kappa_1}{\widetilde{\tau}_1 - \widetilde{\tau}_2 + \kappa_2}}, \qquad \widetilde{\tau}_i = \max(1,\kappa_i) + \widetilde{\mu}_i - \frac{1}{2}|\kappa_i - 1|, \quad i=1,2.
\end{equation}

The obtained advantageous state $\ket{\widetilde{\psi}_{c_{\kappa_1 \mu_1 \kappa_2 \mu_2}}^{AB}}$ has an interesting feature regarding the distribution of energy between the modes ($\frac{1}{1+c^2}$ quanta in the first mode, $\frac{c^2}{1+c^2}$ quanta in the second mode). In Fig.~\ref{figure2}(b) we observe that in the case when the first mode is amplified and the second one is attenuated, sometimes it is beneficial to put more energy in the first (amplified) mode to protect entanglement. Another especially counterintuitive result takes place if both modes are asymmetrically amplified ($\kappa_{1,2} \geq 1$) and $\kappa_2 > \kappa_1 + 2$: then $c_{\kappa_1 \mu_1 \kappa_2 \mu_2} > 1$ irrespective to the noise admixed, i.e., it is always beneficial to put more energy in the second mode with a higher power gain to preserve entanglement.

\section{Conclusions} \label{section-conclusions}

The current study shows that the entanglement of specific low-energy non-Gaussian states has stronger resilience against local Gaussian noises as compared to that of arbitrary-energy Gaussian ones. A similar observation was made earlier in the case of symmetric Gaussian noise $\Phi^A(\kappa,\mu) \otimes \Phi^B(\kappa,\mu)$; however, there was a loophole in the case of asymmetric Gaussian noises $\Phi^A(\kappa_1,\mu_1) \otimes \Phi^B(\kappa_2,\mu_2)$. In this study, we have completely closed this loophole by a rigorous proof on the existence of an entangled non-Gaussian state whose entanglement can tolerate as high noises $\mu_1$ and $\mu_2$ as meet the requitement \eqref{main-inequality}. The Guassian states do lose entanglement beyond the region \eqref{gaussian-inequality}, which is a subset of \eqref{main-inequality}. Moreover, we have explicitly characterized the advantageous non-Gaussian single-photon skewed state $\ket{\widetilde{\psi}_{c_{\kappa_1 \mu_1 \kappa_2 \mu_2}}}$ that exhibits resilient entanglement and discussed peculiarities of the energy distribution among the modes in this state. Our results imply that the optimization problem of maximizing time $t$ during which the state $\Phi^A(\kappa_1(t),\mu_1(t)) \otimes \Phi^B(\kappa_2(t),\mu_2(t))[\ket{\psi^{AB}}\bra{\psi^{AB}}]$ remains entangled definitely has a solution $\ket{\psi^{AB}}$ beyond the set of Gaussian states. Conceptually, these results stimulate further research (initiated in the papers~\cite{sharma-2022,mishra-2022,nair-2018,nair-2022}) beyond the ``Gaussian world'' paradigm in quantum information science (within which solutions to optimization problems involving Gaussian channels and Gaussian measurements are supposed to be attained at Gaussian states and Gaussian ensembles).

\begin{acknowledgments}
The authors are indebted to the referees for bringing Refs.~\cite{adesso-2011} and \cite{lee-2020} to their attention. S.F. is grateful to Mark Wilde for comments on previous works beyond the ``Gaussian world'' paradigm~\cite{sharma-2022,mishra-2022,nair-2018,nair-2022}.
\end{acknowledgments}


\end{document}